# Scaling behavior of the degree of circular polarization of surface plasmon polariton

**Authors**

Dongha Kim[1,2,*], Donghyeong Kim[1,2], Sanghyeok Park[1], N. Asger Mortensen[2,3], and Min-Kyo Seo[1]

**Affiliation**

[1]Department of Physics, KAIST, Daejeon, 34141, Republic of Korea

[2]POLIMA – Center for Polariton-driven Light-Matter Interactions, University of Southern Denmark, Campusvej 55, DK-5230 Odense M, Denmark

[3]Danish Institute for Advanced Study, University of Southern Denmark, Campusvej 55, DK-5230 Odense M, Denmark

*Corresponding Author: dongha@stanford.edu

**Abstract**

Surface plasmon polaritons (SPPs) carry transverse optical spin within the evanescent field, which has enabled the demonstration of various chiral light-matter interactions in classical and quantum systems. To achieve high spin selectivity in the interactions, the elliptical polarization of the evanescent field should be made circular, but the engineering principle of the degree of circular polarization (DOCP) of SPPs has been lacking. In this study, we theoretically and numerically investigate the scaling behavior of the DOCP of the SPP field with respect to the modal effective refractive index ($n_{\text{eff}}$). The DOCP of the SPP field exhibits power-law scalability to the effective refractive index in the 1D layered system, regardless of the material, structural geometry, and excitation wavelength. The power-law scalability is also confirmed in 2D waveguide structures for in-plane and out-of-plane SPP fields, but the scaling exponents vary depending on the distance from the waveguide boundaries by the reduced symmetry of the given system. Due to Lorentz reciprocity, the power-law scalability can be extended to the coupling directionality of chiral emitters towards the plasmonic waveguide. To this end, we propose a chiral photonic platform for enhanced light-valley interaction, which utilizes simultaneous enhancement of the DOCP and coupling directionality. An incident SPP can excite a chiral emitter with high spin selectivity that unidirectionally couples the emitted light into the plasmonic waveguide depending on the valley polarization of excitons in 2D material. Our work provides a ground rule for designing chiral nanophotonic systems and paves the way for the exploration of scale-free phenomena of electromagnetic waves.

**Introduction**

The transverse spin is a fundamental attribute of confined electromagnetic waves, specifically referring to the chiral evanescent field oriented parallel to the direction of propagation [1]. Surface plasmon polaritons (SPPs) offer a pronounced confinement of the evanescent field, allowing for the incorporation of transverse spin in diverse chiral interactions between light and matter [2], such as spin-dependent unidirectional coupling [3], enantioselective sensing [4], spin-assisted optical manipulation [5], and valley-dependent routing of photoluminescence from 2D excitons [6,7]. In the last decade, chiral light-matter interaction in the near-field regime has received significant attention in the context of waveguide quantum electrodynamics [8,9], pursuing high-efficiency coupling between photons and spin qubits in various platforms. However, the purity of chirality, also called the degree of circular polarization (DOCP), which is crucial for the high spin selectivity of light-matter interactions and high signal-to-background ratio detection, has not been fully considered. The recent suggestion of employing elliptical quantum emitters for unidirectional coupling in photonic crystal waveguides has brought attention to the challenge of imperfect chiral fields [10]. In this context, it is necessary to establish an engineering rule of photonic waveguides for near-perfect DOCP to exploit the currently existing quantum emitters with circular polarization basis, such as quantum dots [11], diamond vacancies [12], and van der Waals exciton materials [13]. Introducing an engineering guideline for the DOCP of SPP fields will open new avenues for exploring classical and quantum interactions between chiral light and matter.

In this study, we present a theoretical investigation on the universal scaling behavior of the DOCP of SPP fields regarding to the modal dispersion. In one-dimensional (1D) slab waveguides, we analytically and numerically show that the DOCP exhibits a power-law scalability with the effective refractive index ($n_{eff}$). Regardless of structural parameters, optical material, and excitation wavelength, the DOCP converges to unity (perfect circular polarization) and the power-law scalability solidifies as $n_{eff}$ increases. The power-law scalability is also found for both out-of-plane and in-plane transverse spin in two-dimensional (2D) rectangular waveguides. The scaling exponent of the DOCP can change depending on the transverse location in the rectangular waveguide due to the edge effect, but no pronounced excitation wavelength dependencies is found. Furthermore, we find that the coupling directivity of chiral emitters to a plasmonic waveguide shows the same power-law scalability regarding to $n_{eff}$ due to Lorentz reciprocity. Finally, we propose an experimental scheme of frequency transduction of the SPP using valley-polarized emitters and the engineering rule of unidirectional coupling efficiency.

**Result**

**DOCP of SPP in 1D system**

We begin our discussion of the SPP in an insulator-metal-insulator (IMI) multilayer structure (Figure 1a, left panel). The evanescent field distribution of the lowest-order transverse-magnetic (TM) mode for $z > h/2$ is described by [14],

$$H_y(x, y, z) = A e^{i\beta x} e^{-\kappa_i z}$$

$$E_x(x, y, z) = i \frac{\kappa_i}{\omega \varepsilon_0 \varepsilon_i} A e^{i\beta x} e^{-\kappa_i z}$$

$$E_z(x, y, z) = -\frac{\beta}{\omega \varepsilon_0 \varepsilon_i} A e^{i\beta x} e^{-\kappa_i z} \quad (1)$$

where $\beta = \tilde{n}_{\text{eff}} k_0$ is the propagation constant, $\tilde{n}_{\text{eff}} = n_{\text{eff}} + i k_{\text{eff}}$ is the complex effective refractive index of the given SPP mode, $\kappa_i$ is the attenuation constant ($\kappa_i^2 = \beta^2 - k_0^2 \epsilon_i$), $\omega$ is the angular frequency, $\varepsilon_0$ is the permittivity of the vacuum, $\varepsilon_i$ is the permittivity of the $i$-th insulating layer, and $A$ is a unitless amplitude. The electric fields are further expressible using a circular polarization basis:

$$E_{+\sigma}(x, y, z) = E_z - iE_x = (-\beta + \kappa_i) \frac{A}{\omega \varepsilon_0 \varepsilon_i} e^{i\beta x} e^{-\kappa_i z}$$

$$E_{-\sigma}(x, y, z) = E_z + iE_x = (-\beta - \kappa_i) \frac{A}{\omega \varepsilon_0 \varepsilon_i} e^{i\beta x} e^{-\kappa_i z}$$

where $+\sigma$ ($-\sigma$) corresponds to the right (left) circular polarization. The DOCP is a measure of the visibility between the intensities of the left and right circular polarizations (Figure 1b), which can be defined for any electromagnetic field $E$ lying in a certain 2D plane as

$$\text{DOCP} = \frac{E^* \times E}{|E|^2} = \frac{|E_{+\sigma}|^2 - |E_{-\sigma}|^2}{|E_{+\sigma}|^2 + |E_{-\sigma}|^2}.$$

The DOCP varies from −1 to +1, corresponding to the pure left and right circular polarization, respectively. The DOCP in between −1 and +1 reflects an elliptical polarization state, e.g. 0.5 signifies a superposed state of the left and right circular polarization with a ratio of 1 : 1/3. Given the predominance of the real-part ($n_{\text{eff}}$) of $\tilde{n}_{\text{eff}}$ over the imaginary part ($k_{\text{eff}}$), the DOCP of the SPP field in the $zx$-plane is derived as

$$\text{DOCP} = \frac{2\beta\kappa_i}{\beta^2 + \kappa_i^2} = \frac{\sqrt{1 - \frac{\varepsilon_i}{n_{\text{eff}}^2}}}{1 - \frac{\varepsilon_i}{2n_{\text{eff}}^2}}. \quad (2)$$

Finally, if $n_{\text{eff}}$ is large ($n_{\text{eff}} \gg 1$), the DOCP is approximated as

$$\text{DOCP} \approx 1 - \left(\frac{\varepsilon_i}{2}\right)^2 n_{\text{eff}}^{-4}. \quad (3)$$

Three representative features of the DOCP of the SPP field are found from Eq. (2) and (3). First, the DOCP of the SPP field shows a positive correlation with $n_{\text{eff}}$, which implies that dispersion engineering is the key to the generation of a high-purity chiral evanescent field. We note that a pure circular polarization state (DOCP = ±1) is prohibited for the SPP because it requires the condition of $n_{\text{eff}} \to \infty$, which is the surface plasmon limit, where the plasmon does not propagate. Second, the power-law scalability appeared in the $n_{\text{eff}}$-DOCP relation, which has been commonly reported in the study of universality classes of scale-invariant phenomena [15, 16], such as seismic waves, magnetization reversal, and dynamics of superfluids. We can now estimate the DOCP of the SPP field by $n_{\text{eff}}$, independent of the excitation wavelength, material, and structural parameters of the given plasmonic system. Lastly, the DOCP can vary for the same mode depending on the refractive index of the dielectric background in which the field is located. Low refractive index media are required for high DOCP; for example, the air superstrate yields a higher DOCP than the $SiO_2$ substrate in a $SiO_2$/metal/air configuration.

We first investigated the relation between the DOCP and $n_{\text{eff}}$ of the lowest-order SPP mode in an Air/Ag/Air multilayer employing the finite-difference frequency-domain (FDFD) simulation (Figures 1c-1e). The simulation domain size is 4000 nm × 1 nm with the periodic boundary condition in the lateral spatial axis. The thickness of the Ag slab ($h$) and excitation wavelength ($\lambda$) are scanned from $h$ = 10 to 100 nm and from $\lambda$ = 400 to 800 nm, respectively. The obtained $n_{\text{eff}}$ shows exponential decay tendency with respect to $h$ and $\lambda$ (Figure 1c). The dependence on $h$ comes from the field confinement; the smaller (larger) $h$ results in the stronger (weaker) confinement of the evanescent field in the Ag layer, giving the larger (smaller) attenuation constant κ. The dependence on $\lambda$ is linked to the Drude-like material dispersion of Ag of which the permittivity approaches to the background refractive index as $n_{\text{eff}}$ increases. Meanwhile, the DOCP of the SPP field exhibits a decreasing tendency with $h$ and $\lambda$. The DOCP approaches unity as the Ag slab becomes thinner and the wavelength becomes shorter, which corresponds to the condition for large $n_{\text{eff}}$ (Figure 1d). The $n_{\text{eff}}$-DOCP relation is summarized in Figure 1e and agrees excellently with the asymptotic behavior toward unity DOCP for high $n_{\text{eff}}$ predicted by Eq. (2).

The correlation between the DOCP and effective refractive index can be explained by comparing the amplitudes of the in-plane and out-of-plane electric fields (Figure 1f). Due to the π/2 phase retardation, the amplitude ratio between the $E_x$ and $E_z$ components determines the DOCP. Figure 1f shows the plot of $|E_x|$ and $|E_z|$ depending on $n_{eff}$, normalized to $|E_z|$ at $n_{eff} = 1$, the case of the plane wave. It should be noted that $|E_x|$ and $|E_z|$ are proportional to the attenuation ($\kappa$) and propagation ($\beta$) constants, respectively, following to Eq. (1). For large $n_{eff}$, $\kappa$ becomes approximately same with $\beta$ and thus the $|E_x|$ and $|E_z|$ fields become asymptotic to each other. This tendency can also be understood from the surface charge distribution of the SPP mode (inset). For small (large) $n_{eff}$, the periodicity of the charge oscillation is long (short) so that the lateral electric field becomes small (large) ($E \propto 1/d$, where $d$ is spatial separation of positive and negative charge). Note that the relation between the electric field and the effective refractive index can be extended to a dielectric slab waveguide system due to the identical expression of the field distribution [17]. In other words, the power-law scalability between the DOCP and the effective refractive index is universal for bounded electromagnetic waves of TM polarization.

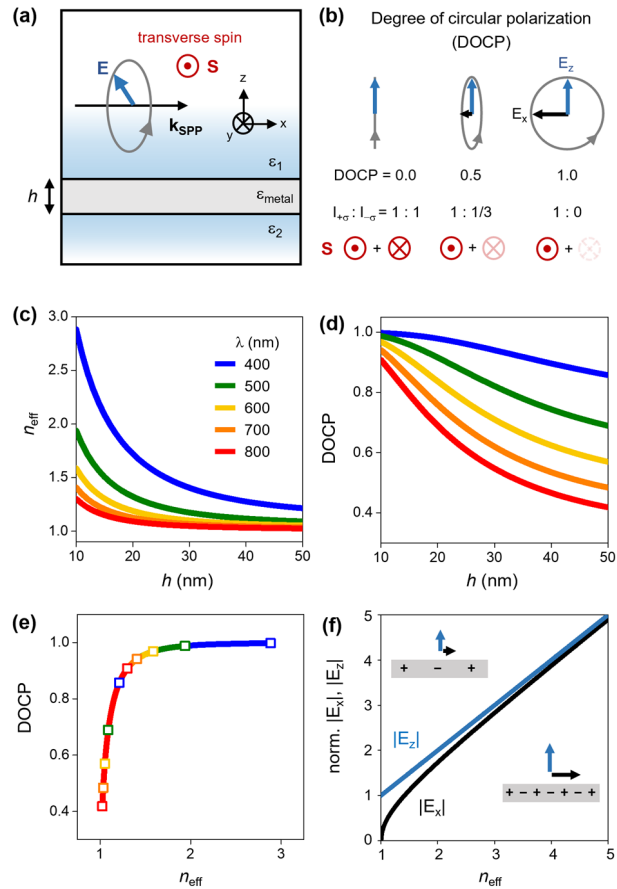

**Figure 1** DOCP of the SPP field in a 1D system. **(a)** Schematic illustration of the electric field distribution (**E**) and transverse spin (**S**) of SPP in an insulator/metal/insulator multilayer structure. **(b)** Description of the DOCP of 0.0, 0.5, and 1.0. The intensities of the left and right circular polarization states are presented as a superposition of the incoming and outgoing transverse spins with respect to the plane of the sheet. **(c, d)** Calculated effective refractive index ($n_{eff}$) and DOCP of the lowest-order SPP mode in the Air/Ag/Air multilayer structure depending on the thickness of the Ag slab ($h$). **(e)** Summary of the relations between $n_{eff}$ and DOCP for the entire dataset. The hollow rectangles indicate the minimum and maximum DOCP for each excitation wavelengths. **(f)** Normalized amplitudes of out-of-plane ($E_z$, blue line) and in-plane ($E_x$, black line) electric field of SPP depending on $n_{eff}$. (inset) Charge distribution and expected electric fields for small and large $n_{eff}$.

We investigate the power-law scalability of the DOCP of SPPs in various layered systems (Figure 2). The $n_{eff}$ and DOCP are collected from 10 acquisition sites (Site 1, 2, 3, ..., 10) in 9 different configurations with materials and structures as described in Figure 2a. For the air/Ag/SiO$_2$ structure, we collected the simulation data from each air superstrate (Site 9) and SiO$_2$ substrate (Site 10). The wavelength ranges to examine the $n_{eff}$ and DOCP are 400-800 nm and 600-800 nm for the structures containing Ag and Au, respectively. The wavelength range for Au is reduced to avoid the absorption by the interband transitions. The slab thickness is scanned from 10 to 100 nm. Figure 2b shows a log-log plot of $n_{eff}$ and 1−DOCP from all acquisition sites with corresponding colors. The plot of 1−DOCP shows a large variation from 0.6855 to 5.988×10$^{-4}$ depending on $n_{eff}$. We set a criterion of the near-perfect circular polarization state as DOCP > 0.9 (blue shaded region), which is the superposed state of ~95% left and ~5% right circular polarization state, and vice versa. The near-perfect circular polarization states are hard to be achieved in the surface modes in semi-infinite metals, while easier in the waveguide modes in metal slabs thanks to the strong field confinement ready to be engineered by the slab thickness. Regardless of its value, the calculated DCOP is in perfect agreement with the analytical estimate of the fourth-order power-law scalability of Eq. (3), as $n_{eff}$ increases (black solid line). For $n_{eff}$ > 2, the value of 1−DOCP exactly follows the power law with the scaling exponent of −4 (gray dashed line). Note that the refractive indices of Ag and Al are extracted from Ref. [18] and the refractive index of SiO$_2$ is set to 1.45.

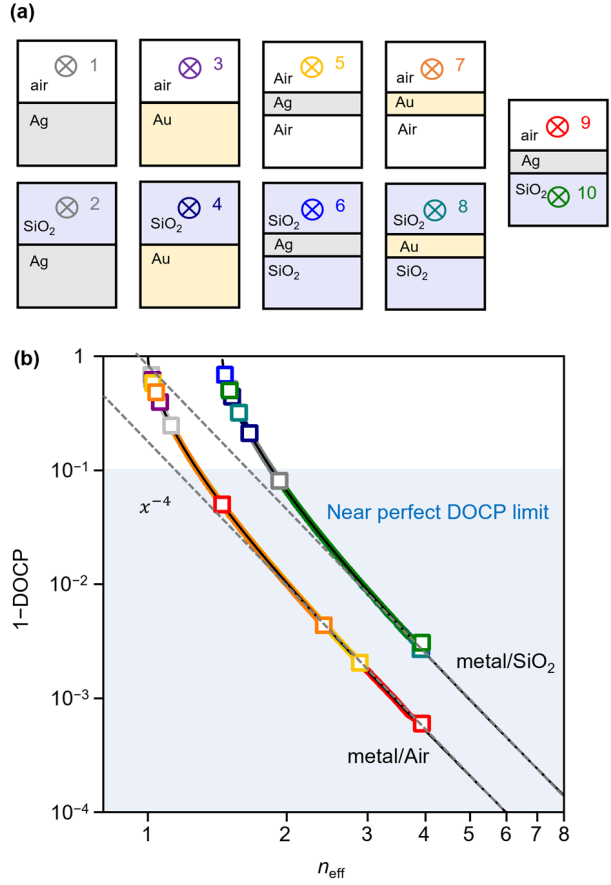

**Figure 2** Power-law scalability of DOCP in a 1D system. **(a)** Schematic illustration of 9 different configurations of materials and structure compositions. The numbers 1 to 9 indicate the site of acquiring the DOCP acquisition. **(b)** Summary of $n_{\text{eff}}$ and (1−DOCP) in a log-log scale. The squares indicate the maximum and minimum of the DOCP for each acquired site. The blue shaded region indicates a near-perfect circular polarization state with a DOCP of >0.9. The black line is the analytic result from Eq. (2). The gray dashed line is the power-law fitting with a scaling exponent of −4.

**DOCP of SPP in 2D system**

We extended our investigation to 2D systems, especially the rectangular plasmonic waveguide on a substrate (Figure 3). Unlike 1D layered systems, the rectangular waveguide-on-substrate configuration is challenging to study analytically due to the reduced symmetry. We conducted FDFD simulations on the Air/Ag rectangular waveguide/$SiO_2$ configuration with systematic variations of the excitation wavelength and the width and height of the waveguide (Figure 3a). The 2D field localization of the SPP propagating along the $x$-axis results in two different transverse spins, the $z$-directional spin normal to the $xy$-plane and the $y$-directional spin normal to the $zx$-plane. The DOCP of each transverse spin, $DOCP_z$ and $DOCP_y$, is given by

$$\text{DOCP}_z = \frac{|E_x + iE_y|^2 - |E_x - iE_y|^2}{|E_x + iE_y|^2 + |E_x - iE_y|^2}$$

$$\text{DOCP}_y = \frac{|E_z + iE_x|^2 - |E_z - iE_x|^2}{|E_z + iE_x|^2 + |E_z - iE_x|^2}.$$

Figure 3b shows the obtained 2D profile of electric field intensity ($|E|^2$), $\text{DOCP}_z$, and $\text{DOCP}_y$ of the lowest-order SPP mode in the cross-sectional plane ($yz$-plane) of the 100 nm wide and 20 nm high Ag waveguide on a $SiO_2$ substrate. The simulation domain size is 1000 nm × 1000 nm with the perfectly matched layer (PML) boundary conditions. The distribution of total electric field intensity is nearly isotropic in the $yz$-plane with strong confinement. The $\text{DOCP}_y$ demonstrates a sign inversion near the air/$SiO_2$ interface, showing positive values in air and negative values in $SiO_2$, and becomes zero due to the absence of an $E_z$ contribution. Likewise, the $\text{DOCP}_z$ switches sign across the center of the waveguide, with positive ($+\hat{y}$) values in the $+y$ region and negative ($-\hat{y}$) values in the $-y$ region, and becomes zero at $y = 0$.

The relation between the DOCP and $n_{\text{eff}}$ is systematically investigated with respect to the width ($w$) and height ($h$) of the rectangular waveguide at the excitation wavelength of 600 nm (Figure 3c). To simplify the analysis, we fixed the acquisition position for the DOCP (the white circles in Figure 3b) at a fixed distance ($l$) of 40 nm from the waveguide boundaries. For $\text{DOCP}_y$, we acquired the values from the air superstrate and the $SiO_2$ substrate, referred to as $\text{DOCP}_{y,1}$ and $\text{DOCP}_{y,2}$, respectively. The obtained $n_{\text{eff}}$ distribution changes from 3.518 for ($w$, $h$) = (20 nm, 20 nm) to 1.640 for ($w$, $h$) = (100 nm, 100 nm). The increasing tendency of $n_{\text{eff}}$ towards small ($w$, $h$) is also resulted from the field confinement as in the 1D system, where the field confinement becomes stronger in a thin and narrow rectangular waveguide. All DOCPs follows the same tendency with $n_{\text{eff}}$; increasing for small ($w$, $h$) and decreasing for large ($w$, $h$). A near-perfect DOCP state appears for all transverse spins, but the $\text{DOCP}_z$ and $\text{DOCP}_{y,1}$ support the state in a broader range of ($w$, $h$) conditions than $\text{DOCP}_{y,2}$. We complied the relation between $n_{\text{eff}}$ and DOCP in Figure 3d for all excitation wavelength ($\lambda$ = 500, 600, and 700 nm) and width and height of the waveguide ($w$, $h$ = 20 – 100 nm, 20 – 100 nm) (Figure 3d). All DOCPs revealed asymptotic converging tendencies toward unity at high $n_{\text{eff}}$, which confirms the universality of the relation between $n_{\text{eff}}$ and DOCP in both 1D and 2D system. The slight deviation in $\text{DOCP}_{y,2}$ distribution can be originated from the field distortion from the substrate, unlike the situation in a 1D systems.

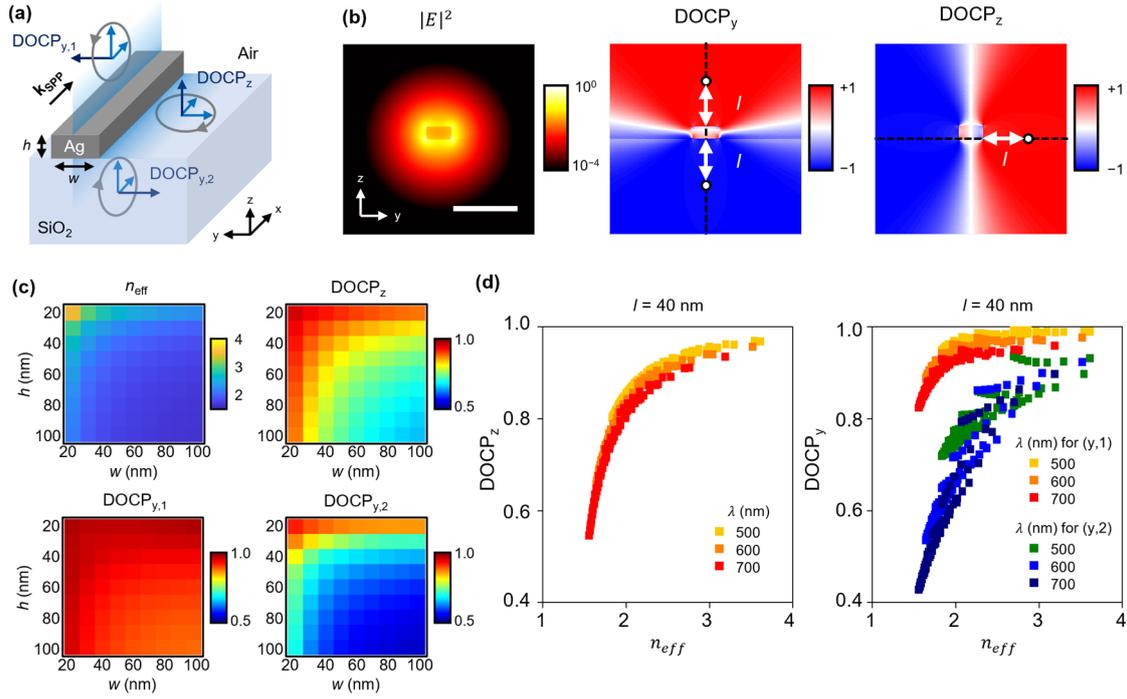

**Figure 3** DOCP of SPP field in 2D system. **(a)** Schematic illustration of the Ag rectangular waveguide on the SiO$_2$ substrate. In-plane and out-of-plane transverse optical spins are analyzed by DOCP$_y$ and DOCP$_z$, respectively. **(b)** Simulated profiles of the total electric field intensity $|E|^2$, DOCP$_y$, and DOCP$_z$ for $(w, h) = (100$ nm, 40 nm). The white circles indicate the acquisition positions of the DOCP at a spatial distance of $l$ from the waveguide boundary. **(c)** Calculated $n_{eff}$, DOCP$_z$, DOCP$_{y,1}$, and DOCP$_{y,2}$ as a function of $w$ and $h$ at $l = 40$ nm. **(d)** Summary of $n_{eff}$ and $(1-\text{DOCP}_z)$ (left) and $(1-\text{DOCP}_y)$ in log-log scale. For DOCP$_z$ and DOCP$_{y,1}$, the excitation wavelengths are indicated in yellow (500 nm), orange (600 nm), and red (700 nm). For DOCP$_{y,2}$, the excitation wavelengths are indicated in green (500 nm), blue (600 nm), and navy (700 nm).

The power-law scalability of the DOCP of the SPP field in 2D systems with respect to $n_{eff}$ was examined, considering different vertical distances from the waveguide ($l$) and various excitation wavelengths ($\lambda$) (Figure 4). The $n_{eff}$ significantly changes with the dimensions of the Ag waveguide, specially its width and height ($w$, $h$), ranging from 20 to 100 nm. We focused on the analysis of the power-law scalability of DOCP$_z$ considering the high application potential of chiral quantum emitters on the substrates, which are further discussed in Figure 5 and Figure 6. Figure 4a shows a log-log plot of the calculated $1-\text{DOCP}_z$ as a function of $n_{eff}$ at $\lambda = 600$ nm. The results clearly demonstrate a power-law scalability of $1-\text{DOCP}_z$ as for $n_{eff} > 2$ in similar to the case of 1D systems. The scaling exponent ($\tau$) changes from 1.498 to 1.842, 2.451, and 3.164 as $l$ increases from 10 to 20, 40, and 100 nm, respectively. Figure 4b presents a comprehensive examination of the behavior of $\tau$ as a function of $d$ at different excitation wavelengths of 500, 600, and 700 nm. It is worth noting that $\tau$ exhibits a consistent pattern across all the wavelengths examined, revealing universality of the scaling behavior. As $l$ increases from

5 to 200 nm, the value of τ varies from ~1 to ~4.

The position-dependent nature of the power-law scaling arises from the fact that the evanescent electric field of the two-dimensional SPP waveguide does not follow a simple, isotropic exponential decay in space. Figure 4c shows the logarithmic plot of the amplitude of the $x$- and $y$-components of the electric field for two representative waveguides with $(w, h) = $ (20 nm, 20 nm) and (100 nm, 100 nm). Unlike the logarithm of a simple exponential function, $\log(|E_x|)$ and $\log(|E_y|)$ show nonlinear behavior, changing their slopes, $\kappa_x$ and $\kappa_y$, as a function of $l$. In addition, $\kappa_x$ and $\kappa_y$ differ from each other, which implies that the relative contributions of the $E_x$ and $E_y$ fields to the $DOCP_z$ change with distance from the waveguide. As the difference between $|E_x|$ and $|E_y|$ decreases for high $l$, the magnitude of $DOCP_z$ increases as shown in Figure 3b. In contrast, a 1D system features symmetric confinement of the $E_x$ and $E_y$ fields along the out-of-plane direction, resulting in identical values of $\kappa_x$ and $\kappa_y$ and eliminating polarization dependence. To delve deeper, we calculate $\kappa_x$ and $\kappa_y$ and plotted the ratio $\kappa_x/\kappa_y$ against $n_{eff}$ (Figure 4d). The extracted $\kappa_x/\kappa_y$ varies with $l$ but they approach to unity for high $n_{eff}$. Hence, we expect that the power-law scalability is maintained by the $n_{eff}$ independency in $\kappa_x/\kappa_y$, but the scaling exponent can be changed depending on $\kappa_x/\kappa_y$. At $l = 100$ nm, $\kappa_x/\kappa_y$ approaches to unity, which converges to the case of a 1D system with the scaling exponent of 4.

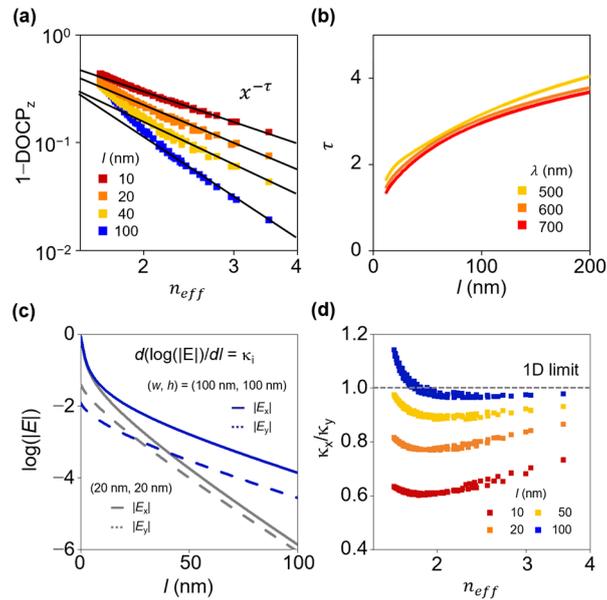

**Figure 4** Power-law scalability of SPP in 2D system. **(a)** Log-log plot of $1-DOCP_z$ versus $n_{eff}$ for various distances $d$: 10 (red), 20 (orange), 40 (yellow), and 100 (blue) nm. The solid black lines are the fitted power-law scaling. **(b)** Extracted scaling exponent (τ) depending on $d$ for the excitation wavelengths of 500, 600, and 700 nm, respectively. **(c)** Plot of $\log|E_x|$ (solid line) and $\log|E_y|$ (dashed line) as a function of $d$ for two representative Ag waveguides with $(w, h) = $ (20 nm, 20 nm) (gray) and (100 nm, 100 nm) (blue), respectively.

(d) Plot of the ratio between $\kappa_x$ and $\kappa_y$, the spatial derivative of $\log|E_x|$ and $\log|E_y|$, versus $n_{eff}$ for different $d$ of 10 (red), 20 (orange), 40 (yellow), and 100 (blue) nm. The gray dashed line indicates the condition of $\kappa_x = \kappa_y$, which is the case for a 1D system.

**Scaling behavior of coupling directionality of chiral emitter**

We explore the engineering principle of the coupling directionality of a chiral emitter towards a plasmonic waveguide (Figure 5). It is well known that the reciprocity theorem between an oscillating current and the resulting field reveals that in any photonic system, the external quantum efficiency of a quantum emitter and the field localization of incident light are correlated [19]. Likewise, the coupling directionality of a chiral emitter to the SPP waveguide corresponds to the reciprocal counterpart of the DOCP (Figure 5a). The coupling directionality is predominantly governed by the transverse spin of the evanescent field, leading to an asymmetric coupling of the light emitted by a chiral emitter, which can be modeled as a circularly polarized dipole, to the SPP modes propagating in different directions [5, 6, 9, 10]. For instance, a chiral emitter with $-z$-directional circular polarization directs its emission mainly to the SPP mode propagating in the $+y$ direction, with a small fraction coupling to the $-y$ direction. The ellipticity of the evanescent field results in the simultaneous coupling of $\pm y$ propagating modes, and unidirectional coupling can be demonstrated when the magnitude of the DOCP of the SPP field is close to unity. The coupling directionality is defined as $C = (I_+ - I_-)/(I_+ + I_-)$, where $I_+$ ($I_-$) is the intensity of the $+y$ ($-y$) propagating SPP mode coupled from the chiral emitter. Employing the finite-difference time-domain (FDTD) simulation, we calculate $C$ of the chiral emitter coupled to the Ag rectangular waveguide. The chiral emitter is located on the air/SiO$_2$ interface at a distance of $l$ from the boundary of the waveguide. Figure 5b is the simulated electric field intensity profile of the chiral emitter near the 20 nm wide and 20 nm high Ag waveguide. The calculated $C$ is 0.87 and 0.98 for $l = 10$ nm and 100 nm, respectively. To calculate $C$, we employ $I_+$ and $I_-$ at the positions $\pm 1.5$ μm away from the chiral emitter from the line plot of the electric field intensity distribution in Figure 5c.

We reveal that $C$ exhibits identical features to those we observed for the DOCP$_z$. Figure 5d is the calculated $C$ depending on the width and height of the Ag waveguide at $d = 40$ nm and $\lambda = 600$ nm. The distribution of $C$ mirrors that of the DOCP$_z$ in Figure 3c. Indeed, due to their reciprocal relation, the DOCP$_z$ and $C$ have a perfect positive correlation, DOCP$_z$ = $C$, for all structural parameters (Figure 5e). It has to be obvious that $C$ adheres the power-law scalability of the DOCP$_z$ and the position dependence of the scaling exponent. Figure 5f shows the power-law scalability of $1-C$ as a function of $n_{eff}$ and the scaling exponents depending on the distance of the chiral emitter from the waveguide precisely align with those observed for the $1-$DOCP$_z$: $\tau = 1.477, 1.829, 2.265$, and $3.079$ for $d = 10, 20, 40$, and $100$ nm, respectively. In addition, the log-log plot of $1-C$ for different excitation wavelengths

(500, 600, and 700 nm) shows no spectral dependence, as shown in Figure 5g. Consequently, the dispersion engineering of $n_{eff}$ improves not only the DOCP of the SPP field but also boosts the spin-selectivity of chiral light-matter interactions.

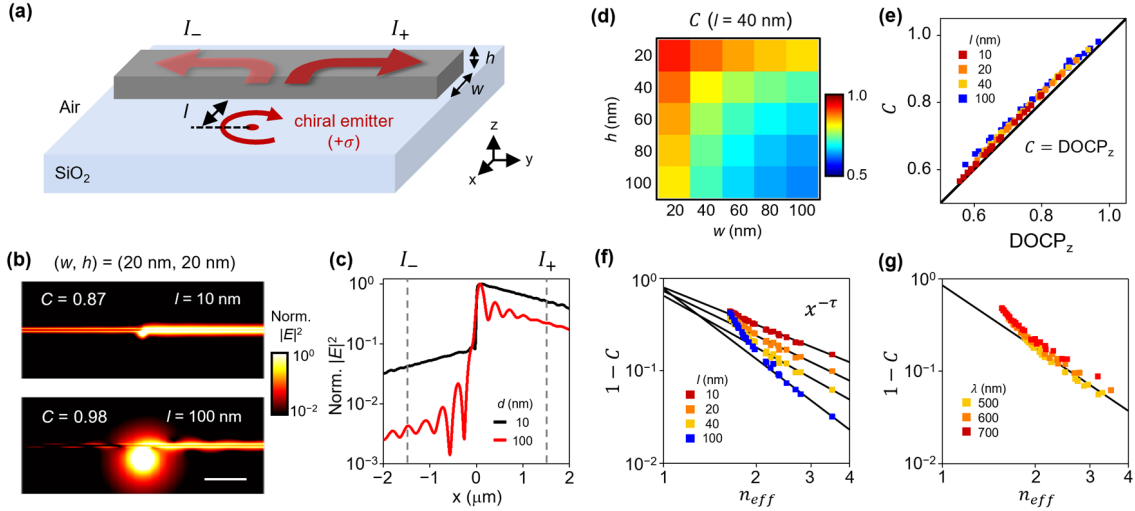

**Figure 5** Correlation between the coupling directionality and DOCP. **(a)** Schematic illustration of the Ag rectangular waveguide and the chiral emitter with the spatial separation $l$. The intensity of the coupled SPP propagating in the $+y$ ($-y$) direction is expressed as $I_+$ ($I_-$). **(b)** Simulated total electric field intensity profile of propagating coupled SPP modes depending on $l = 10$ and $100$ nm for Ag waveguide with $(w, h) = (20$ nm, $20$ nm). Scale bar is 500 nm. **(c)** Line plot of normalized electric field $|E|^2$ distribution for $l$ (nm) = 10 (black), 100 (red), respectively. For the calculation of coupling directionality $C$, the intensities for each $+y$ and $-y$ propagating SPP were extracted 1.5 μm away from the chiral emitter. **(d)** Calculated C distribution depending on $w$ and $h$ at $l = 40$ nm. **(e)** Plot of $DOCP_z$ and $C$ for $l = 10$ (red), 20 (orange), 40 (yellow), and 100 nm (blue). **(f, g)** Analysis of the power-law scalability of $1-C$ with respect to $n_{eff}$ in a log-log plot: **(f)** for spatial separations $l$ at 10, 20, 40, and 100 nm, and **(g)** for excitation wavelengths λ of 500, 600, and 700 nm.

**Enhanced light-valley interaction in chiral photonic platform**

Finally, we propose a chiral photonic platform for enhanced light-valley interactions, utilizing the simultaneous enhancement of the DOCP and the coupling directionality in a plasmonic waveguide system. Figure 6a illustrates the experimental scheme of frequency transduction and routing of SPPs in an Ag waveguide combined with a high valley-polarization 2D material, such as multilayer $WS_2$ [5,6,20] or various moiré heterostructures [21,22]. When the $DOCP_z$ is near-perfect under high $n_{eff}$, the chiral field of the incident pump SPPs (λ = 594 nm) can selectively excite 2D excitons at a specific valley (+K or −K) in the band structure of the multilayer $WSe_2$. The excited excitons can serve as chiral quantum emitters with valley-polarization $\rho = (I_{+\sigma} - I_{-\sigma})/(I_{+\sigma} + I_{-\sigma})$, where $I_{\pm\sigma}$ represents the intensity of emitted photoluminescence with $\pm\sigma$ polarization under $+\sigma$ excitation. If $\rho$ is $\pm 1$, the emitter emits photoluminescence of a perfect chirality with the $\pm\sigma$ polarization state, which couples back to the propagating SPPs ($\lambda = 620$ nm) of the Ag waveguide. Consequently, the emitted photoluminescence unidirectionally couple to waveguide due to near-unity $C$, where the propagation direction is determined by the sign of $\rho$. We note that the frequency transduction indicates the frequency difference between the incident and the directionally coupled SPP.

The performance of the proposed platform is numerically investigated by integrating the FDFD and FDTD simulation results (Figures 6b and 6c). We estimate the coupling directionality of the photoluminescence of 2D excitons of the multilayer WSe$_2$ excited by, and in turn emitting to, the SPP mode of the Ag waveguide. As a proof of concept, we simplified the distribution of 2D excitons as a linear chain of circularly polarized dipoles along the direction ($x$-axis) normal to the waveguide. The normalized intensities of $+\sigma$ and $-\sigma$ polarizations in the pumping SPP field at a certain distance ($l$) are $I_{e,\pm\sigma}(l) = I_e(l)(1 \pm \mathrm{DOCP}_z(l))/2$, where $I_e(l)$ is the intensity of in-plane electric field. For a given valley polarization $\rho$, the intensities of excited chiral emitters at $l$ are derived as,

$$I_{q,\pm\sigma}(l) = \frac{1 \pm \rho}{2} I_{e,\pm\sigma} + \frac{1 \mp \rho}{2} I_{e,\mp\sigma}$$

The resulting electric field intensity profile ($I_f$) of coupled SPP and coupling directionality ($C_{2D}$) are derived as,

$$I_f(x,y) = \int dl \left( I_{q,+\sigma}(l) I_{C,+\sigma}(x,y;l) + I_{q,-\sigma}(l) I_{C,-\sigma}(x,y;l) \right) \tag{4}$$

$$C_{2D} = \frac{I_{f,+y} - I_{f,-y}}{I_{f,+y} + I_{f,-y}} \tag{5}$$

where $I_{f,\pm y} = I_f(x = 0 \text{ μm}, y = \pm 1.5 \text{ μm})$ are the intensity of coupled SPP at 1.5 μm away from the 2D excitons. In the integration of $I_f(x, y)$, we ignored the presence of the 2D material beneath the Ag waveguide due to the quenching effect. The length of the 2D material is set to 500 nm, which is sufficiently long compared to the decay length of the evanescent electric field of ~100 nm in Figure 3b.

We numerically demonstrate the frequency transduction and routing of the SPP propagation depending on the valley-polarization of the 2D material. Figures 6b show the resulting electric field intensity profile ($I_f$) of the SPP depending on the height of the Ag waveguide ($h$) and the valley polarization ($\rho$) based on Eq. (4) and (5). Here, we fix the width of the waveguide as $w = 100$ nm. The propagation of the coupled SPP can be maneuvered depending on $\rho$ as unidirectional toward the same ($+y$, $\rho = 1$) or the reverse ($-y$, $\rho = -1$) direction compared to incident SPP, and omnidirectional ($\rho = 0$).

The unidirectionality can be optimized for small $h$ where the intensity of the coupled SPP propagating in the opposite direction ($-y$) decreases. Figure 6c and 6d are the summary of the resultant coupling directionality $C_{2D}$ depending on $h$ and $\rho$. Following the tendencies of the DOCP and the $C$, the resultant coupling directionality $C_{2D}$ increases for small $h$, where $C_{2D}$ is 0.6415 at $h$ = 20 nm as 0.6415 (Figure 6c). The $C_{2D}$ is linearly proportional to $\rho$, which implies that valley-polarization of the 2D material not only improves the unidirectionality but also determines the propagation direction of the coupled SPP. We expect the improvement of $C_{2D}$ can be further achieved in the ultra-thin, ultra-narrow plasmonic waveguide ($w$, $h$ < 20 nm) on the area of 2D material with the injecting/outcoupling tapered waveguide.

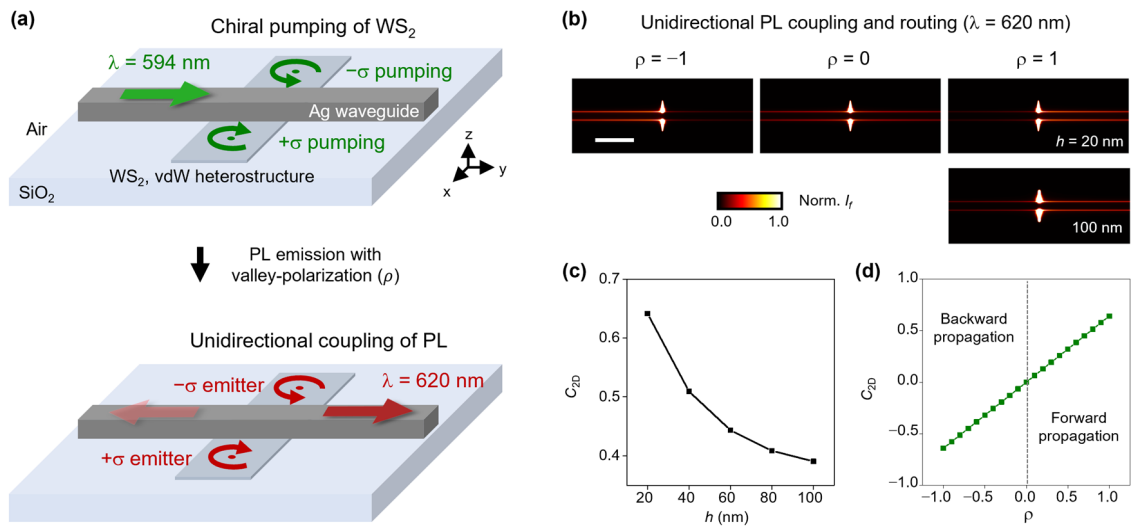

**Figure 6** Proposed chiral photonic platform with enhanced light-valley interaction. **(a)** Working principle of given platform that utilizes simultaneous enhancement of DOCP and $C$. (upper) Excitation of chiral quantum emitters by incident SPP ($\lambda$ = 594 nm) with high DOCP$_z$ evanescent field. The DOCP of photoluminescence (PL) from chiral emitters are governed by valley-polarization ($\rho$) of given 2D material. (lower) Emitted light from the chiral emitters directionally couples to SPP ($\lambda$ = 620 nm) by high $C$. **(b)** Resultant electric field intensity ($I_f$) profile of coupled SPP from chiral emitters depending on $h$ and $\rho$. Scale bar is 500 nm. **(c,d)** Calculated coupling directionality of chiral emitters in 2D material ($C_{2D}$) depending on **(c)** $h$ and **(d)** $\rho$.

## Discussion

We investigated an engineering rule for the DOCP of the evanescent field based on dispersion engineering of the SPP. From 1D to 2D plasmonic systems, power-law scalability of 1−DOCP is numerically confirmed with respect to $n_{eff}$, the simulation data being in good agreement with the analytic results from 1D system. We also showed power-law scalability of the coupling directionality of chiral emitters towards the SPP mode, enabling us to propose a chiral photonic platform with enhanced light-valley interaction. Given that power-law scaling behavior often suggests universality across various physical implementations, we anticipate the extension of our discoveries to other waveguide systems, facilitating high spin-selective interactions between light and matter, such as on-chip waveguides [23,24] and optical fibers [25,26]. Regarding to the dispersion engineering, atomically thin optical waveguides are intrinsically advantageous to achieve near-perfect DOCP states in the evanescent fields due to the giant $n_{\text{eff}}$, such as graphene, transition metal-dichalcogenides, and 1D atomic arrays [27,28,29,30]. These constitute promising platforms for recent waveguide quantum electrodynamics, implying high-spin selectivity in the coherent control of qubits. Finally, our exploration of the scaling behavior of the DOCP charts a new course for investigating universality classes in electromagnetic waves. This involves examining diverse power-law scaling behaviors in the phases of light as it traverses certain critical phenomena.

**Data Availability**

The data that support the findings of this study are available from the corresponding author upon reasonable request.

**Acknowledgement**

M.-K.S. acknowledges the support of the National Research Foundation of Korea (NRF) (2020R1A2C2014685). D.K. acknowledges the support of the NRF (2019R1A6A1A10073887). The Center for Polariton-driven Light-Matter Interactions (POLIMA) is funded by the Danish National Research Foundation (Project No.~DNRF165).

**Author Contribution**

Dongha Kim (D. K.), Donghyeong Kim and N. A. M. conceived the idea. D. K. conducted analytic and numerical calculations. D. K. and N. A. M. analyzed the data. D. K., N. A. M., and M.-K. S. wrote the manuscript with input from all other authors.